\documentclass[12pt]{extarticle}
\usepackage{graphicx}
\usepackage{framed}
\usepackage[normalem]{ulem}
\usepackage{amsmath}
\usepackage{amsthm}
\usepackage{amssymb}
\usepackage{amsfonts}
\usepackage{enumerate}
\usepackage[utf8]{inputenc}
\usepackage[top=1 in,bottom=1in, left=1 in, right=1 in]{geometry}
\usepackage{indentfirst}
\usepackage{float}
\usepackage[pdftex]{hyperref}
\usepackage [boxruled, linesnumbered]{algorithm2e}

\theoremstyle{definition}

\title{COVID - 19: A model for studying the evolution of contamination in Brazil}
\author{ Rodrigo André Schulz$^1$ \\ Carlos H. Coimbra-Araújo$^{1,2}$ \\ Samuel Willian Schwertner Costiche$^1$\footnote{samuel.costiche@ufpr.br}}
\date{
{\small $^1$Departamento de Engenharias e Exatas - Universidade Federal do Paraná, Pioneiro, 2153, 85950-000, Palotina, PR, Brazil.}\\
{\small $^2$Programa de Pós-Graduação em Física Aplicada - Universidade Federal da Integração Latino Americana, 85866-000, Foz do Iguaçu, PR, Brazil.}\\
\vspace{1cm}
\today}

\begin{document}

\maketitle
\begin{abstract}
In the present article we introduce an epidemiological model for the investigation of the spread of epidemics caused by viruses. The model is applied specifically to COVID-19, the disease caused by the SARS-Cov-2 virus (aka ``novel coronavirus''). The SIR (Susceptible - Infectious - Recovered) model is used as a basis for studying the evolution of the epidemic. Nevertheless, we have modified some of the model hypotheses in order to obtain an estimate of the contamination free of overestimated predictions. This extended model is then applied to the case of the recent advance of the epidemic in Brazil. In this regard, it is possible to obtain the evolution for the number of infectious significantly close to that provided by current data. Accordingly, we evaluate possible future scenarios for the disease spread. Regarding the population susceptibility, we consider different social behaviors in response to quarantine measures and precautions to avoid contagion. We conclude that the future scenario of the epidemic depends significantly on the social behavior adopted to date, as well as on the contagion control measures. The extent of such measures would be likely to cause thousands, millions or tens of millions of contaminations in the next few months.
\end{abstract}


\section{Introduction}
The emergence of SARS-CoV-2 epidemic around the world has motivated a series of studies and projections for the evolution of the disease over the next few months. The fight against the spread of the disease occurs in the world by the use of many research techniques, treatments and the prevention of the contamination\ \cite{1}. In the latter case, one of the main factors associated with virus prevention is related to restrictions in human social contact in order to prevent the unrestrained contagion\ \cite{2}.

However, the absence of a broad scientific literature on the evolution of the virus, as well as the accelerated growth of contamination in Brazil during March 2020, give rise to a series of possible predictions. When such predictions are disclosed by the main information vehicles, it is provided scope for minimalist and maximalistic interpretations of the case. In this aspect,  adequate projections based on current data is needed.

Considering such circumstances, it is natural that epidemiological models arise driven by actual available data related to the evolution of the disease. Once created the model, it is possible to predict possible future scenarios from which one can have a better indication of the dimension of such an epidemic in the country and in the world.

Some of the currently known models used in such procedures are SIS, SIR and SIRS\ \cite{3}. The first one is used mainly in the study of diseases in which recovery does not prevent the re-contamination of the pacient, usually caused by bacterial agents\ \cite{4}. The second is used for modeling epidemics involving infectious diseases, such as COVID-19, caused by SARS-COV-2\ \cite{5}. The third one is basically used for the study of infections caused by influenza, since it allows the modeling of a situation where recovered individuals lose their immunity (which SIR, for example, does not model)\ \cite{6}. 

The present work explores the building of a variation of the SIR model in order to cover relevant conditions present in the Brazilian context, such as: 1) daily mortality and daily birth rates (which change, over time, the population) and 2) the gradual reduction of the population susceptible to the disease in fuction of social distancing measures. Here it will be considered a modification in the hypothesis that, initially, the entire population analyzed is susceptible to contamination by the disease.

The Brazilian situation until March 22 had more than 1,500 confirmed cases of infection, with 25 cases of death from the disease\ \cite {7}. The exponential growth of infection cases reveals the need to develop studies related to the behavior of epidemics in order to stabilize the current scenario, as well as to allow the advance in the development of tools that permit analyzing the behavior of epidemics in the future from the first cases. Other relevant impact of such studies is that they provide concrete justifications for the effectiveness and awareness related to social distancing policies. The paper is divided in sections. The second section reviews the original formulation of the SIR model. The third section presents the methodology and the forth section deals with the use of such a methodology in the Brazilian case. The fifth and the final sections present some discussion, concluding remarks and prospects.

\section{SIR Model: the original formulation}
The SIR model is based in a simple hypothesis: the individual of a given population where an epidemic occurs goes through different stages of susceptibility to infection\ \cite{3}. Such stages give rise to well defined compartments in the model, where the individual is classified as:
\begin{center}
    Susceptible
    
    Infectious
    
    Recovered
\end{center}
From this hypothesis, an epidemic can be characterized as a flow:
\[S\rightarrow I\rightarrow R,\]
where the infectious $I$ and recovered $R$ population grow over time, while the susceptible $S$ subject decreases over time. In this way, this flow can be described by functions $S(t)$, $I(t)$ and $R(t)$ such that the evolution of $I(t)$ over time characterizes the number of cases of infection at all times throughout the course of the disease.

To characterize the temporal evolution of the model, it is necessary to establish how the instantaneous variation of the functions of the model occurs. For example, if $\alpha$ is the rate of change in the number of infectious individuals (the ratio of infectious to a previous time interval), then the number of new infections at each time interval, in a population of $N$ individuals, reduces the susceptible population as follows:
\begin{equation}
    \frac{dS(t)}{dt}=\frac{-\alpha I(t)S(t)}{N}.\
    \label{1}
\end{equation}

\noindent On the other hand, if $\beta$ is the rate of recovery from the epidemic, then the number of recovered individuals will be a fraction of the number of infectious:
\begin{equation}
    \frac{dR(t)}{dt}=\beta I(t).
    \label{2}
\end{equation}

\noindent At last, the number of infected each day should equal the difference between the number of new contaminations in the susceptible population and the number of recovered from the infected population, that is
\begin{equation}
    \frac{dI(t)}{dt}=\frac{\alpha I(t)S(t)}{N}-\beta I(t).
    \label{3}
\end{equation}
The SIR model also considers that the initial value associated with the functions $S(t)$, $I(t)$ and $R(t)$ can be defined assuming that, initially, the entire population is susceptible to infection, that there is a minimum population $d$, initially infectious (otherwise there would be no way for the epidemic to start), and there is no one recovered, since the epidemic did not started. In other words:
\[S(0) = N,\ I(0)=d,\ R(0)=0.\]
In this way, the temporal evolution of the functions described by the SIR model consists of an initial value problem involving a system of three ordinary differential equations.

\section{Review of the model and methodology}

In the previous section, the SIR model was presented as a proposal for modeling an epidemic. The system of equations is reasonably simple. However, its construction requires some hypotheses whose acceptance may imply an overestimated forecast of the number of people infected by the epidemic. Are they:
\newline
\newline
I - Initially, the entire population is susceptible to contamination.
\newline
\newline
II - There are no deaths or births over the course of the epidemic.
\newline
\newline
III - There are no reductions in the susceptible population, for example, due to quarantine measures.
\newline

Those processes significantly interfere in estimating the number of people infected. In respect to I, for example, why should one suppose that the contamination of one person in São Paulo on February 26 would imply that someone in the state of Roraima ($\sim 4,500$ km away) would be susceptible to contamination on February 26? The susceptibility to contamination is directly related to the proximity between contaminated individuals, but in this case, if both are separated by long distances, without maintaining any contact, then there is no reason to suppose that there is a relationship of susceptibility between them. Regarding II, it is well known that several deaths of people, contaminated or not by the disease, occur throughout its evolution, interfering in the number of susceptible individuals. Similarly, the more people born the more susceptible to contamination they become, promoting an increase in the number of cases. And in respect to III, it is assumed that, with the evolution of the epidemic, people will begin to isolate themselves socially, whether by individual will or governmental determination, so that the susceptible population is also reduced due to this factor\ \cite{8}.

This means that the forecast of the SIR model is naturally overestimated to calculate the number of people infected daily on a value of $S(t)$ which can be many times greater than the real one.
\subsection{Alterations in the model}
To correct the problematic points in assertions I, II and III, one can add terms in the equations (\ref{1}), (\ref{2}) and (\ref{3}) in order to operate them according to the logic of a growing susceptible population, where there are reductions resulting from deaths and from social isolation processes, as well as an increase in the number of susceptible people due to the birth rate.
\subsubsection{Gradual increase in the susceptible population}

Given a population $N$, the SIR model originally supposes that $S(0) = N$ and $S(t)<S(0), \forall t>0$. Here, in order to correct the fact pointed out in I, it is supposed that $S(0) = (1-\rho_{0}) N$, where $\rho_{0}$ is the percentage of the population initially isolated from contamination, and that $\frac{dS(t)}{dt}$ evolves so that the rate $\rho_{0}$ decreases over time, i.e.
\begin{equation}
    \frac{dS(t)}{dt}=\frac{-\alpha I(t)S(t)(1-\rho(t))}{N},
    \label{4}
\end{equation}

\noindent where 
\begin{equation}
   \rho (t) = \rho_{0}(1-\Delta \rho)^t,
   \label{5}
\end{equation}
\noindent and $\Delta \rho$ is the mean rate of the growing of the susceptible population with the disease evolution. In this respect, the smaller the value of $\rho_{0}$ is (a value between 0 and 1) the more the initial population approaches $N$. Besides, the greater $\Delta \rho$ is the faster the susceptible population approaches the total population. For example, a reasonable estimate for $\rho_{0}$ is to consider the initial susceptible population as the population of the place where the epidemic begins, taking the $\Delta \rho$ rate as the mean percentage of the growing of the susceptible population, parameterized by the total time required for the epidemic to reach all the domains of a given country. That is, if $\Delta t*$ is the time required for the epidemic to reach all the domains (states) of a country in which the epidemics occurs, then the mean number of new susceptible individuals in each day will be
\[n = \frac{N}{\Delta t*},\]
and the percentage of new susceptible people each day will be
\[\Delta \rho = \frac{n}{N}.\]
In this case, $\Delta \rho$ can also be interpreted as the inverse of the period necessary for the epidemic to reach the whole country, because
\begin{equation}
    \Delta \rho = \frac{n}{N}=\frac{\frac{N}{\Delta t*}}{N}=\frac{1}{\Delta t*}.
    \label{6}
\end{equation}

This change also implies that the number $I (t)$ of contaminated individuals will increase at the same rate that $S(t)$ decreases, that is, depending on the gradual increase in the susceptible population:
\begin{equation}
    \frac{dI(t)}{dt}=\frac{\alpha I(t)S(t)(1-\rho(t))}{N}-\beta I(t).
    \label{7}
\end{equation}

\subsubsection{Mortality rate and birth rate}
To correct the value of the population susceptible to contamination due to deaths and daily births, just add terms to the equation (\ref{4}) that relate the mortality rate and the daily birth rate to the susceptible population $S(t)$. That is, if $\gamma$ is the percentage of the population that dies daily due to something unrelated to the epidemic, and $\theta$ is the daily rate of people that born in the place where the epidemic occurs, then the variation in the susceptible population will be proportional to the difference between the number of people who born and the number of people who die, i.e.
\[P(t) = (\gamma - \theta) S(t),\]
where $P(t)$ expresses the variation in the number of susceptible people, per day, due to mortality and birth rates. When adding $P(t)$ to the equation (\ref{4}), we get
\begin{equation}
    \frac{dS(t)}{dt}=\frac{-\alpha I(t)S(t)(1-\rho(t))}{N}-P(t).
    \label{8}
\end{equation}

\subsubsection{Population in quarantine}

It may happen that, after some period after the beginning of the evolution of the epidemic, there is a dramatic reduction in the population $S (t)$ due to social distancing measures. The cancellation of classes in public schools and universities, commercial and industrial activities, as well as musical concerts and similar events are examples of how these reductions can occur. Thus, it is possible that there is a time $\tau$ such that the susceptible population will be reduced with the rate $k$ of its default value. When this occurs, the equation (\ref{8}) should start to consider the new population as the contingent under which the contamination factor $\alpha$ acts. To characterize this process, it is possible to define the functions $S (t)$ and $I (t)$ by parts, as follows:
\[\left\{\begin{matrix}
\frac{dS(t)}{dt}=\frac{-\alpha I(t)S(t)(1-\rho(t))}{N}-P(t),\ \mathrm{if}\ t<\tau;\\ 
\\
\frac{dS(t)}{dt}=\frac{-\alpha I(t)S(t)(1-\rho(t)-k)}{N}-P(t),\ \mathrm{if}\ t\geq \tau;\\
\\
\frac{dI(t)}{dt}=\frac{\alpha I(t)S(t)(1-\rho(t))}{N}-\beta I(t),\ \mathrm{if}\ t<\tau;\\
\\
\frac{dI(t)}{dt}=\frac{\alpha I(t)S(t)(1-\rho(t)-k)}{N}-\beta I(t),\ \mathrm{if}\ t\geq \tau.\\
\end{matrix}\right.\]

This process occurs in such a way that the population $S(t)$ tends to increase with time, since $\rho(t)$ decreases with time, but it also tends to decrease with the factor $k$ (from $\tau$). That is, if the proportion $k$ of people who begin to isolate themselves in quarantine is greater than the rate at which the susceptible population increases (parameterized by $\rho(t)$), then the epidemic will begin to decrease. On the contrary, if the rate at which people become susceptible is greater than the rate at which they become quarantined, then the epidemic process will continue to grow until it reaches its maximum.

It is also possible to generalize this definition by parts so that, for each $t_{1}, t_{2}, t_{3}, ..., t_{n}$ where quarantine processes are started, one can respectively consider the factors $k_{\tau}$,  $k_{\tau + \Delta t_{1}}$, $k_{\tau + \Delta t_{2}}, ..., k_{\tau + \Delta t_{n}}$ which act by decreasing the susceptible population. Therefore, from the $n$th instant $t$, the equations $\frac{dS(t)}{dt}$ and $\frac{dI (t)}{dt}$ will be considered under the form:

\[\left\{\begin{matrix}
\frac{dS(t)}{dt}=\frac{-\alpha I(t)S(t)(1-\rho(t))}{N}-P(t),\ \mathrm{if}\ t<t_{n};\\ 
\\
\frac{dS(t)}{dt}=\frac{-\alpha I(t)S(t)(1-\rho(t)-k_{\tau}-k_{\tau+\Delta t_{1}}-k_{\tau+\Delta t_{2}}-...-k_{\tau+\Delta t_{n}})}{N}-P(t),\ \mathrm{if}\ t\geq t_{n}>\tau;\\
\\
\frac{dI(t)}{dt}=\frac{\alpha I(t)S(t)(1-\rho(t))}{N}-\beta I(t),\ \mathrm{if}\ t<t_{n};\\
\\
\frac{dI(t)}{dt}=\frac{\alpha I(t)S(t)(1-\rho(t)-k_{\tau}-k_{\tau+\Delta t_{1}}-k_{\tau+\Delta t_{2}}-...-k_{\tau+\Delta t_{n}})}{N}-\beta I(t),\ \mathrm{if}\ t\geq t_{n}>\tau.\\
\end{matrix}\right.\]

\subsection{Discretization}

The obtained system of equations, after the proposed modifications, can be summarized as follows:
\[\left\{\begin{matrix}
\frac{dS(t)}{dt}=\frac{-\alpha I(t)S(t)(1-\rho(t)-k_{\tau}-k_{\tau+\Delta t_{1}}-k_{\tau+\Delta t_{2}}-...-k_{\tau+\Delta t_{n}})}{N}-P(t),\ t\geq t_{n}>\tau;\\ 
\\
\frac{dI(t)}{dt}=\frac{\alpha I(t)S(t)(1-\rho(t)-k_{\tau}-k_{\tau+\Delta t_{1}}-k_{\tau+\Delta t_{2}}-...-k_{\tau+\Delta t_{n}})}{N}-\beta I(t),\ t\geq t_{n}>\tau;\\ 
\\
\frac{dR(t)}{dt}=\beta I(t);\\
\\
S(0) = N;\ 0<\rho _{0}<1;\ I(0)=0;\ R(0)=0;\ \alpha,\ \beta, \ \gamma, \Delta \rho,\ k_{\tau+\Delta t_{n}}  \in \mathbb{R}.  \\

\end{matrix}\right.\]

A very simple way to find the solutions to the initial value problem is to obtain an approximation for the derivatives of $S(t)$, $I(t)$ and $R(t)$ from the Taylor Series and, following the Euler method, obtain the values of $S(t+\Delta t)$, $I(t+\Delta t)$ and $R(t+\Delta t)$ as follows:
\[S(t+\Delta t) = S(t) + \frac{dS(t)}{dt}\Delta t,\]
\[I(t+\Delta t) = I(t) + \frac{dI(t)}{dt}\Delta t,\]
\[R(t+\Delta t) = R(t) + \frac{dR(t)}{dt}\Delta t,\]
where $\Delta t$ is the step of the solution, determined by the upper and lower limits of the interval at which the solution is calculated\footnote{See the Appendix}, as well as the number of subdivisions in the $n$ range, that is:
\[\Delta t =\frac{T_{f}-T_{i}}{n},\]
so that the solution is calculated in the interval $[T_{i}, T_{f}]$.

\section{The case of Brazil}
In view of the reformulation of the SIR model, we can consider its application to describe the evolution of the COVID-19 epidemic. In Brazil, it can be considered that it started on February 26, 2020\ \cite{2,7,12}. First, it is necessary to characterize the parameters that define the evolution of the epidemic in the country. The Brazilian population is among the largest in the world, with around $211,720,000$ people\ \cite{9}. Thus, for simulating the evolution of the epidemic in Brazil, it can be assumed that:
\[N=211,720,000\ \mathrm{people}.\]

In addition, for the simulation, it is assumed that the mortality rate in the country follows the global annual rate of $7.7/1,000$ people per year\ \cite{10}. In order to simulate a daily evolution, the daily percentage rate corresponding to the number of deaths per year must be considered:
\[\gamma = \frac{7.7}{1,000}\times\frac{1}{365\  \mathrm{days}}=\frac{2.11\times 10^{-5}}{\mathrm{day}}.\]

Regarding the birth rate, Brazil registered 2,899,851 births in 2018, according to data from IBGE (Brazilian Institute of Geography and Statistics) \cite{11}. In this way, the mean daily birth rate, defined in terms of the 2018 data, provides a daily birth rate given by

\[\theta = \frac{2,899,851}{211,720,000}\times \frac{1}{365\ \mathrm{days}}\cong \frac{3.75\times 10^{-5}}{\mathrm{day}}.\]

According to information released by the Adolfo Lutz Institute, the first confirmed case of COVID-19 in Brazil occurred on February 25, 2020, in the case of a patient who was in São Paulo\ \cite{12}. Thus, since it is necessary to assume a minimum initial value of contamination to evolve the epidemic, the simulation will occur from February 26th to the current date. Therefore, being sure of the first contamination, and the active action of the contaminated person in relation to the spread of the virus, for the simulation, it is assumed that
\[d = 1,\ T_{i} =0 \rightarrow\ \mathrm{February}\ 26.\]

As the patient was in the state of São Paulo, it is reasonable to assume that the population initially susceptible is that of the state itself. Thus, it is assumed that $\rho_{0}$ (the population initially free from contamination) is the total population, except for the population of the state of São Paulo, about 21 \% of the Brazilian population\ \cite{13}. So:
\[\rho_{0} = 1-0.21 = 0.79.\]
In addition, until March 21 (24 days after the outbreak of the epidemic), the Ministry of Health recognized that community transmission of the virus had reached the entire country\ \cite{14}, so that the period required for susceptible population to reach the entire country is 24 days, and thus, from equation (\ref{6}):
\[\Delta \rho = \frac{1}{24}\cong 0.0417.\]

Finally, it remains to determine the values of $\alpha$ and $\beta$ from the current scenario of the epidemic in Brazil. According to data obtained by the Worldometer\ \cite{15} platform, Brazil had, until March 28, the number of 3,904 confirmed cases of COVID-19 distributed in the time series illustrated in Figure 1.
\begin{figure}[H]
    \centering
    \includegraphics[scale = 0.7]{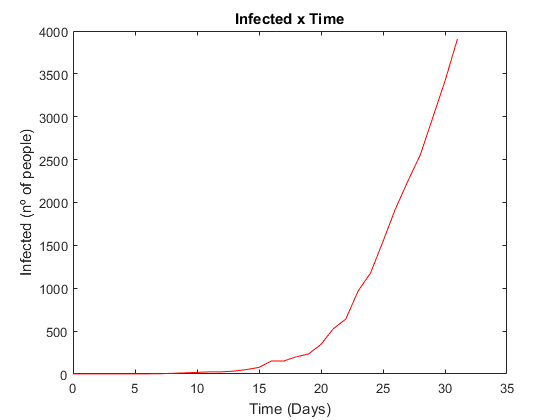}
    \caption{COVID-19 real contamination in Brazil. Source: Worldometer\ \cite{15}.}
    \label{fig:my_label}
\end{figure}

The mean growth rate $\alpha$ is defined as the mean of the ratios of the number of cases on the following day in relation to the previous day. That is, in this case:
\[\alpha = \frac{1}{25}\sum_{n=1}^{25} \frac{I(T_{i}+n)}{I(T_{i}+n-1)}\cong 1.39,\]
where $n = 1$ corresponds to February 26, and $n = 25$ corresponds to March 22\footnote{The average based on this period provides the contamination factor for the moment when there were still no social isolation processes that strongly interfere in the progress of the disease. Using a long period to calculate the mean growth could result in a value that differs significantly in the initial evolution of the epidemic.}. For the recovery factor $\beta$, a similar procedure is performed, considering the mean of the ratio of the number of recovered to the number of contamination cases. Due to the unavailability of the data, however, it is assumed that the Brazilian recovery factor follows the world average\ \cite{14}, i.e.
\[\beta=\frac{1}{25}\sum_{n=1}^{25} \frac{R(T_{i}+n)}{I(T_{i}+n-1)}\cong 0.32.\]

Finally, it is also assumed that, starting on March 14, there are five factors \footnote{The value of $k_{n}$ indicates the $k$ factor that acts in the susceptible population from the $n$th day. As the epidemic started on February 26, March 14 corresponds to the 17th day of the epidemic. Same with the others.} ($k_{17}$, $k_{22}$, $k_{23}$, $k_{26}$, and $k_{28}$) that make it to be reduced by 20 \%, 5 \%, 5 \%, 10 \%, and more 5 \% on March 14, March 19, March 20, March 23, and March 25, respectively. These values are arbitrary due to the lack of concrete data about how many people are in fact isolated from contamination by quarantine, and express the trend, from these dates, of the reduction of contamination determined by the variations in the the curve illustrated in Figure 1.

Considering these data, the result of the simulation between February 26 and March 28 is illustrated in Figure 2.
\begin{figure}[H]
    \centering
    \includegraphics[scale = 0.6]{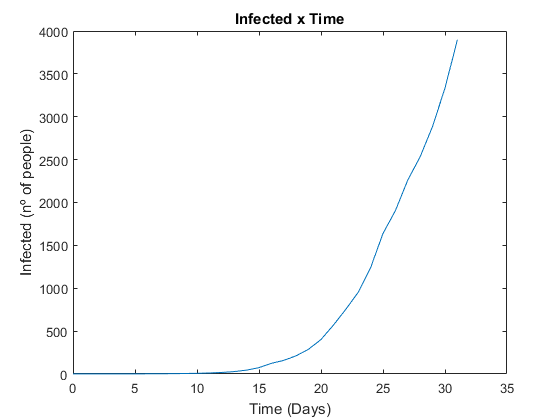}
    \caption{Simulation of the COVID-19 evolution in Brazil.}
    \label{fig:my_label}
\end{figure}
Figure 3 shows the comparison between the curves in Figures 1 and 2.
\begin{figure}[H]
    \centering
    \includegraphics[scale = 0.7]{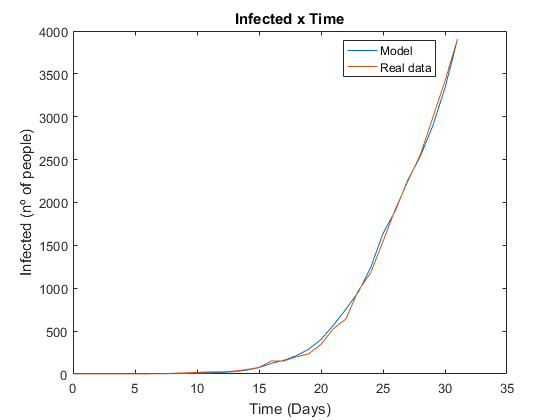}
    \caption{Comparison between simulation and real data for the evolution of COVID-19 in Brazil.}
    \label{fig:my_label}
\end{figure}

Given the satisfactory adjustment close to the curve of real data, one can consider the extent of the simulation results for the future. However, as there is currently no specific scenario regarding the factors of reduction of the susceptible population due to social distancing measures, several possibilities can be explored. The first of these is not to consider, from April onwards, that social isolation measures are taken, that is, to remove the $k_{31}$ factor from the 34th day of the disease evolution (corresponding to the 31st of March).

Two other possible cases can be explored by imposing conditions of future social distancing, in order to verify the effect of such measures. Figure 4 shows the evolution of the epidemic for the first case, where there are no restrictive measures in the future, and two other cases where the population promotes social isolation keeping it at 50 \% until the end of the epidemic and, lastly, raising the percentage of the quarantined population by 20 \% from April 10, that is, the comparison between the absence of social isolation measures, and measures reaching 50 \% and 70 \% of the population, respectively

\begin{figure}[H]
    \centering
    \includegraphics[scale = 0.7]{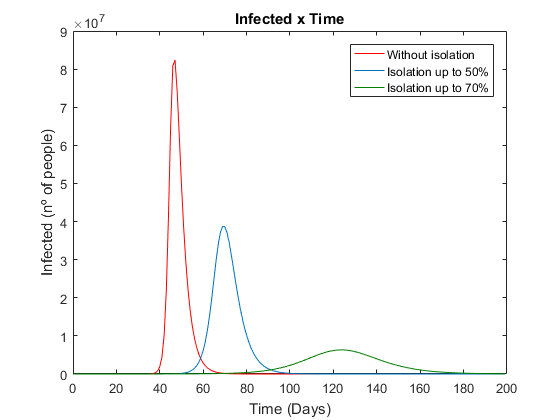}
    \caption{Comparison between the evolution of the COVID-19 epidemic in Brazil for 3 possible cases.}
    \label{fig:my_label}
\end{figure}

In the first case, a worrying scenario occurs, where the peak of contamination reaches practically 39 \% of the population on April 13. In the second case, the 50 \% reduction in the susceptibility of the population results in a peak of contamination that reaches around 18.3 \% of the Brazilian population on May 5, which consists of a considerable reduction in relation to the first case. In the latter case, with isolation conditions reaching 70 \% of the population, this peak of cases reaches about 3 \% of the Brazilian population, on June 29. 

These results show how social distancing measures alter the dynamics of the epidemic in the country in order to decrease the peak and the total number of cases, as well as to extend the time necessary for the epidemic to reach a maximum in the number of contaminations.

Additionally, these results may raise the following question: given an instant $\tau$, how restrictive should social isolation measures be? That is, what is the best estimate for the value of $k_{\tau}$? In section 3.1.3 we showed how it is possible to consider population reduction rates as a result of social isolation measures. So that, for the instant $\tau$, it follows
\[\frac{dI(\tau)}{dt}=\frac{\alpha I(\tau)S(\tau)(1-\rho(\tau)-k_{\tau})}{N}-\beta I(\tau).\]

\noindent One way to estimate the value of $k_{\tau}$ in this case is to make the number of new contaminated individuals null, i.e.

\[\frac{\alpha I(\tau)S(\tau)(1-\rho(\tau)-k_{\tau})}{N}=0,\]
and therefore
\[(1-\rho(\tau)-k_{\tau})=0,\]

\begin{equation}
    k_{\tau}=1-\rho(\tau)=1-\rho_{0}(1-\Delta \rho)^{\tau}.
    \label{9}
\end{equation}

This condition does not guarantee that new contamination will cease to occur completely, since it would be necessary to keep the entire infected population in isolation, which is not taken into account by the model, and may often not be a measurable data from the country's statistics. In theory, if all people infected by the epidemic are known, then it would not be necessary that the social isolation condition to extend to the susceptible population. However, in cases where an emerging epidemic begins to evolve and the actual number of infected people is not known, the equation (\ref{9}) is the best estimate for how restrictive measures of social isolation should be.

In the case of Brazil, taking March 25 (28 days after the epidemic started), March 31 (34 days after the epidemic started) and April 10 (44 days after the epidemic started) as a reference, we obtain respectively the values:
\[k_{28}=1-0.79\times(1-0.0417)^{28}\cong0.76,\]
\[k_{34}=1-0.79\times(1-0.0417)^{34}\cong0.814,\]
\[k_{44}=1-0.79\times(1-0.0417)^{44}\cong0.88.\]

Thus, it is possible to note that the longer the time at which social isolation measures are implemented, the more restrictive they must be in relation to the number of people quarantined. Thus, given a very long period of time, the value of $k_{\tau}$ will require that the entire population establish measures of social isolation, since
\[\lim_{\tau \rightarrow \infty } k_{\tau }=\lim_{\tau \rightarrow \infty }1-\rho_{0}(1-\Delta \rho )^{\tau }=1,\]

\noindent because $\Delta \rho$ is always greater than zero.
\section{Discussion}
These three possible future scenarios are consistent with the actual situation of the epidemic in Brazil. That is, with respect to the actual data available to date, the three scenarios are possible and depend exclusively on how future actions related to social behavior will be taken.

In this respect, the development of epidemiological models capable of predicting the evolution of the epidemic in the face of social behavior is an important tool for raising awareness, shedding some light on many aspects of the disease pattern and providing strength to the scientific dissemination about the importance of the social responsibility involved in quarantine measures. After all, apparently, although other epidemics may arise in the future, more or less severe than that caused by SARS-CoV-2, the results of such natural phenomena, in terms of the number of infected and dead individuals, depend essentially on human behavior and the priority given to the processes of reducing susceptibility.
\section{Concluding remarks}
The results obtained from modeling the SARS-COV-2 epidemic evolution in Brazil allow one to draw up estimates and predictions for the future scenario of the disease progress in the country. As discussed, both scenarios -- the worrying and the controlable -- are possible and compatible with the current data. In this context, dissemination of scientific facts and awareness-raising campaigns are of paramount worthness, shedding some light on the importance of isolation policies during epidemic situations.

In this article, we reformulated the SIR model to comply with the hypotheses of a susceptible population that grows over time and varies with mortality and birth rates. As well as being able to model susceptibility reductions in function of social measures for controling the epidemic advance. We explored the case of the epidemic evolution in Brazil, a worth issue since it could mean, given the characteristics of the country, a considerable impact on the global economy, not to mention the serious consequences to the country's economic and social structure.

We conclude that, within the social and economic possibilities of the country, it is prudent to foster the maintenance of quarantine policies in order to avoid mass contamination of the population in April, 2020. Nevertheless, the daily updating of data linked to the current situation of the country must be perpetuated, since, for many reasons, it is possible that, to the date, the number of cases are underestimated and, thus, the forecasts provided here may change significantly due to new updates of the real situation in the country.

\begin{center}
    \textbf{Appendix}
\end{center}
The algorithm used to solve the system of ODE's was formulated in Mat Lab language, and it is configured as follows:

\begin{verbatim}
%Algorithm: COVID-19 in Brazil

%Domain

 Tf = input ('Enter the ending time value: ');
 Ti = input ('Enter the start time value: ');
 n = input ('Enter the number of subdivisions of the time range: ');
 dT = (Tf-Ti)/n;
 T = Ti:dT:Tf;

% Constants

alpha = input ('Enter the average value of the contamination rate: ');
gamma = input ('Enter the current death rate value: ')
beta = input ('Enter the average recovery rate: ');
d = input ('Enter the number of initial patients: ');
p(1) = input ('Enter the starting value of the percentage of the population
free from contamination: ');
deltap = input ('Enter the value of the percentage reduction of the
contamination-free population per day: ');
N = input ('Enter the population value: ');
k(1) = input ('Enter the period until 20% of the population is quarantined: ');
k(2) = input ('Enter the period until 25% of the population is quarantined: ');
k(3) = input ('Enter the period until 30% of the population is quarantined: ');
k(4) = input ('Enter the period until 40% of the population is quarantined: ');
k(5) = input ('Enter the period until 45% of the population is quarantined: ');

%Initial conditions

 S(1) = N;
 R(1) = 0;
 I(1) = d;

 %Evolution of the susceptible population
 for i=1:n
     p(i+1) = p(1)*power((1-deltap),(Ti + i*dT))
     if (p(i+1) < 0.1)
         p(i+1) = 0
     end
 end
 
 %Iteration
 
for i=1:n
    if (i< round(k(1)/dT))
        Sder(i) = ((-alpha*I(i)*(S(i)*(1-p(i))))/N)-((gamma-theta)*S(i));
        Rder(i) = beta*I(i);
        Ider(i) = (((alpha*I(i))*(S(i)*(1-p(i))))/N)-(beta*I(i));
        R(i+1) = R(i) + Rder(i)*dT;
        I(i+1) = I(i) + Ider(i)*dT;
        S(i+1) = (S(i) + Sder(i)*dT);
    else if (i > round(k(1)/dT)) && (i  < round(k(2)/dT))
            Sder(i) = ((-alpha*I(i)*(S(i)*(1-p(i)-0.2)))/N)-((gamma-theta)*S(i));
            Rder(i) = beta*I(i);
            Ider(i) = (((alpha*I(i))*(S(i)*(1-p(i)-0.2)))/N)-(beta*I(i));
            R(i+1) = R(i) + Rder(i)*dT;
            I(i+1) = I(i) + Ider(i)*dT;
            S(i+1) = (S(i) + Sder(i)*dT);
        else if (i > round(k(2)/dT)) && (i  < round(k(3)/dT))
                Sder(i) = ((-alpha*I(i)*(S(i)*(1-p(i)-0.25)))/N)-((gamma-theta)*S(i));
                Rder(i) = beta*I(i);
                Ider(i) = (((alpha*I(i))*(S(i)*(1-p(i)-0.25)))/N)-(beta*I(i));
                R(i+1) = R(i) + Rder(i)*dT;
                I(i+1) = I(i) + Ider(i)*dT;
                S(i+1) = (S(i) + Sder(i)*dT);
            else if (i > round(k(3)/dT)) && (i  < round(k(4)/dT))
                    Sder(i) = ((-alpha*I(i)*(S(i)*(1-p(i)-0.3)))/N)-
                    ((gamma-theta)*S(i));
                    Rder(i) = beta*I(i);
                    Ider(i) = (((alpha*I(i))*(S(i)*(1-p(i)-0.3)))/N)-(beta*I(i));
                    R(i+1) = R(i) + Rder(i)*dT;
                    I(i+1) = I(i) + Ider(i)*dT;
                    S(i+1) = (S(i) + Sder(i)*dT);
                else if (i > round(k(4)/dT)) && (i  < round(k(5)/dT))
                        Sder(i) = ((-alpha*I(i)*(S(i)*(1-p(i)-0.4)))/N)-
                        ((gamma-theta)*S(i));
                        Rder(i) = beta*I(i);
                        Ider(i) = (((alpha*I(i))*(S(i)*(1-p(i)-0.4)))/N)-(beta*I(i));
                        R(i+1) = R(i) + Rder(i)*dT;
                        I(i+1) = I(i) + Ider(i)*dT;
                        S(i+1) = (S(i) + Sder(i)*dT);
            else 
                Sder(i) = ((-alpha*I(i)*(S(i)*(1-p(i)-0.45)))/N)-((gamma-theta)*S(i));
                Rder(i) = beta*I(i);
                Ider(i) = (((alpha*I(i))*(S(i)*(1-p(i)-0.45)))/N)-(beta*I(i));
                R(i+1) = R(i) + Rder(i)*dT;
                I(i+1) = I(i) + Ider(i)*dT;
                S(i+1) = (S(i) + Sder(i)*dT);
                    end
                end
            end
        end
    end
end

% Graphics
T_in_days = Ti:1:Tf-1

I_per_day(1) = I(1)
for i=2:Tf
    I_per_day(i) = I(i*(1/dT))
end
plot (T_in_days,I_per_day)
title ('Infected x Time')
xlabel ('Time (Days)')
ylabel ('Infected (nº of people)')

S_per_day (1) = S(1)
for i=1:Tf
    if (i< round(k(1)/dT))
        S_per_day(i) = S(i*(1/dT))*(1-p(i*(1/dT)))
    else if (i > round(k(1)/dT)) && (i  < round(k(2)/dT))
           S_per_day(i) = S(i*(1/dT))*(1-p(i*(1/dT))-0.2) 
        else if (i > round(k(2)/dT)) && (i  < round(k(3)/dT))
               S_per_day(i) = S(i*(1/dT))*(1-p(i*(1/dT))-0.25)
            else if (i > round(k(3)/dT)) && (i  < round(k(4)/dT))
                   S_per_day(i) = S(i*(1/dT))*(1-p(i*(1/dT))-0.30)
                else if (i > round(k(4)/dT)) && (i  < round(k(5)/dT))
                      S_per_day(i) = S(i*(1/dT))*(1-p(i*(1/dT))-0.40)  
                    else 
                        S_per_day(i) = S(i*(1/dT))*(1-p(i*(1/dT))-0.45)
                end
            end
        end
    end
end

plot (T_in_days,S_per_day)
title ('Susceptible x Time')
xlabel ('Time (Days)')
ylabel ('Susceptible people(nº of people)

R_per_day(1) = R(1)

for i=1:Tf
    R_per_day(i) = R(i*(1/dT))
end
plot (T,R)
title ('Recovered x Time')
xlabel ('Time (Days)')
ylabel ('Recovered people (nº of people)')

\end{verbatim}

\end{document}